\documentclass[reprint,superscriptaddress,preprintnumbers,amsmath,amssymb,aps,prd,tightenlines,longbibliography]{revtex4-2}

\usepackage{graphicx}
\usepackage{xcolor}
\usepackage[sort&compress]{natbib}
\usepackage{amsmath,amssymb,bm,bbm,slashed,amsfonts}
\usepackage{xr-hyper}
\usepackage[colorlinks=true
,urlcolor=blue
,anchorcolor=blue
,citecolor=blue
,filecolor=blue
,linkcolor=red
,menucolor=blue
,linktocpage=true
,pdfproducer=medialab
,pdfa=true
]{hyperref}
\usepackage{cleveref}
\usepackage{enumerate}
\usepackage{epsfig, subfigure}
\usepackage{setspace}
\usepackage{booktabs, tabularx}
\usepackage{units}
\usepackage{placeins}
\usepackage{multirow}
\usepackage{mathtools}
\usepackage[normalem]{ulem}

\begin{document}

\title{
Cosmic Birefringence and Axiogenesis
}

\author{Raymond T.~Co}
\affiliation{Physics Department, Indiana University, Bloomington, IN 47405, USA}
\author{Lawrence J.~Hall}
\affiliation{Leinweber Institute for Theoretical Physics, Department of Physics, University of California, Berkeley, CA 94720, USA}
\affiliation{Theoretical Physics Group, Lawrence Berkeley National Laboratory, Berkeley, California 94720, USA}
\author{Keisuke Harigaya}
\affiliation{Enrico Fermi Institute, Kavli Institute for Cosmological Physics, Leinweber Institute for Theoretical Physics, Department of Physics, The University of Chicago, Chicago, IL 60637, USA}
\affiliation{Kavli Institute for the Physics and Mathematics of the Universe (WPI),
The University of Tokyo Institutes for Advanced Study,
The University of Tokyo, Kashiwa, Chiba 277-8583, Japan}
\author{Taegyu Lee}
\affiliation{Physics Department, Indiana University, Bloomington, IN 47405, USA}

\begin{abstract}
Recent analyses of Planck and ACT CMB data report a $2\text{--}4\sigma$ hint of isotropic cosmic birefringence ($\beta \approx 0.3^\circ$), corresponding to a parity-violating rotation of polarization planes that points to potential new physics. We show that both the hint of cosmic birefringence and the observed baryon asymmetry can be simultaneously explained by the rotation of an axion-like field. The minimal setup predicts a positive birefringence angle, consistent with the CMB hint, and matter and kination domination ending around a temperature of $\mathcal O(3\text{--}50)$ MeV, making the model falsifiable by astrophysical probes and potentially giving rise to characteristic gravitational-wave signals. We also introduce a mechanism that can exponentially enhance the angular momentum of the rotation generated by the Affleck-Dine mechanism.
\end{abstract}

\maketitle

\noindent
{\bf Introduction.}---%
Several analyses of CMB polarization have reported a preference for a nonzero isotropic cosmic birefringence angle \cite{Minami:2020odp,Diego-Palazuelos:2022dsq,Eskilt:2022wav,Eskilt:2022cff,Diego-Palazuelos:2022cnh,AtacamaCosmologyTelescope:2025blo,Remazeilles:2025wzd,Diego-Palazuelos:2025dmh,Eskilt:2026imm}.  Planck-based analyses typically find $\beta\simeq0.3^\circ$ with nominal significances of $2$--$4\sigma$, while ACT DR6 gives a consistent independent preference.  A recent joint ACT DR6--Planck DR4 analysis finds~\cite{Eskilt:2026imm}
\[
\beta=0.277^\circ\pm0.057^\circ=(4.83\pm0.99)\times10^{-3}\ {\rm rad},
\]
corresponding to a nominal significance of $4.8\sigma$, reduced to about $3.5\sigma$ in a more conservative foreground-mitigated analysis.  Overall, the hinted angle lies roughly in the range $0.002$--$0.009$ radians. Although instrumental calibration and Galactic foreground systematics remain important, the non-zero angle might be a manifestation of new physics.

Isotropic cosmic birefringence can be explained by a coherent motion of an axion-like particle (ALP) field that couples to the photon~\cite{Carroll:1989vb,Carroll:1991zs,Harari:1992ea,Lue:1998mq,Carroll:1998zi,Liu:2006uh,Finelli:2008jv,Fedderke:2019ajk,Fujita:2020aqt}.  Motion of an ALP field driven by a potential has been considered, and the non-zero ALP energy is identified with (fractional) dark matter or dark energy~\cite{Fujita:2020ecn,Nakagawa:2021nme,Choi:2021aze,Komatsu:2022nvu,Gasparotto:2022uqo,Murai:2022zur}. On the other hand, ALP cosmic strings cannot explain the hint of cosmic birefringence 
because of an excessive anisotropy of cosmic birefringence~\cite{Amin:2026dvh}.

In this {\it Letter}, we demonstrate that {\it rotation} of an ALP can also explain cosmic birefringence. Rotation of the ALP in field space may be initiated in the early universe, continue down to when CMB photons decouple from the thermal bath, and induce cosmic birefringence.

We also show that the baryon asymmetry of the universe can be explained by the same rotation of the ALP via axiogenesis, where the angular momentum of the ALP is partially transferred to the chiral asymmetry of the Standard Model (SM) particles, which is subsequently converted to a baryon asymmetry via the electroweak sphaleron process~\cite{Co:2019wyp,Domcke:2020kcp,Co:2020xlh}.

The energy density of an ALP rotation, $\rho_\theta$, depends on the potential of the radial direction of the complex field in which the ALP resides. For supersymmetric models, the radial potential is (nearly) quadratic. The scaling of $\rho_\theta$ is matter-like $\propto R^{-3}$ (kination-like, $\propto R^{-6}$) when the rotation is on the body (minimum) of the radial potential, where $R$ is the scale factor of the universe~\cite{Co:2019wyp}. Matter-dominated and kination-dominated eras can exist and modify the spectra of pre-existing gravitational waves, e.g., from inflation or cosmic strings~\cite{Co:2021lkc,Gouttenoire:2021wzu,Gouttenoire:2021jhk}.

We find that the angular momentum of the ALP rotation to explain cosmic birefringence is large. We show that such a large angular momentum can be obtained with the aid of a generalization of the Kim-Nilles-Peloso mechanism~\cite{Kim:2004rp,Harigaya:2014eta,Choi:2014rja,Higaki:2014pja,Harigaya:2014rga,Choi:2015fiu}, a.k.a.~the clockwork mechanism~\cite{Kaplan:2015fuy}.

\vspace{0.1cm}
\noindent
{\bf ALP rotation and Cosmic Birefringence.}---%
We consider an ALP $\phi$ coupled to the electromagnetic field via
\begin{equation}
    {\cal L} = \frac{\alpha_{\rm EM}}{8 \pi} c_\gamma \frac{\phi}{f} F_{\mu\nu} \tilde F^{\mu\nu} = - \frac{\alpha_{\rm EM}}{2 \pi} c_\gamma \frac{\phi}{f}{\bf E}\cdot {\bf B},
\end{equation}
where $\alpha_{\rm EM}$ is the fine-structure constant, $c_\gamma$ is a constant, and $f$ is the decay constant. We also define $g_{\phi \gamma \gamma}\equiv \alpha_{\rm EM} c_\gamma/2 \pi f$ and $\theta \equiv \phi/f$.

We assume that the axion field rotates in field space in the early universe, which can be initiated by the Affleck-Dine mechanism~\cite{Affleck:1984fy}. To compute the CMB birefringence angle, we define 
\begin{align}
  Y_\theta \equiv \frac{n_\theta}{s} = \frac{\dot{\theta} f^2}{s},
    \label{eq:thetadot}
\end{align}
where $n_\theta$ and $s$ are the charge and entropy density, respectively. After the initiation process, since both $n_\theta$ and $s$ decrease as $R^{-3}$,  $Y_\theta$ is conserved.

The birefringence angle induced by the ALP is determined by the change of the field value between the last scattering surface (LSS) and today~\cite{Komatsu:2022nvu}
\begin{equation}
    \beta =- \frac{g_{\phi\gamma\gamma}}{2}(\phi_0-
    \langle\phi\rangle_{\rm LSS}) = -\frac{\alpha_{\rm EM} c_\gamma}{4\pi}(\theta_0-
    \langle\theta\rangle_{\rm LSS}).
\end{equation}
The finite thickness of the LSS can be included using the visibility function $\mathcal{V}(T)$~\cite{Fujita:2020ecn},
\begin{equation}
    \langle\theta\rangle_{\rm LSS} = \int dT\;\mathcal{V}(T)\theta(T).
\end{equation}
Using Eq.~(\ref{eq:thetadot}),
\begin{equation}
    \theta_0-
    \langle\theta\rangle_{\rm LSS} = \frac{Y_\theta}{f^2}\int dT\;\mathcal{V}(T)\int^{t_0}_{t(T)}dt' s(t'),
\end{equation}
and we find
\begin{align}
\beta =  0.003\times c_\gamma \left( \frac{-Y_\theta /f^2}{8.7\times 10^{-9}\;{\rm GeV}^{-2}} \right).
\end{align}

The required $Y_\theta$ is
\begin{align}
\label{eq:Yreq}
    Y_\theta= -5.7\times10^{10}c_\gamma\left(\frac{\beta}{0.005}\right)\left(\frac{f/c_\gamma}{2\times10^9\;{\rm GeV}}\right)^2.
\end{align}
The strongest astrophysical constraint on the ALP-photon coupling at masses $m_\phi < 10^{-12}~{\rm eV}$ is~\cite{Reynes:2021bpe}
\begin{align}
g_{\phi \gamma\gamma} = \frac{\alpha_{\rm EM} c_\gamma}{2\pi f} < 6.3 \times 10^{-13}\;{\rm GeV}^{-1},
\end{align}
which requires $f/c_\gamma \gtrsim 2\times 10^{9}$ GeV, implying a large $|Y_\theta| \gtrsim 10^{10} c_\gamma$ from Eq.~\eqref{eq:Yreq}.

\vspace{0.1cm}
\noindent
{\bf Initiation of the rotation.}---%
We discuss how the required $Y_\theta$ can be obtained by the Affleck-Dine mechanism, taking into account the necessary thermalization.

The ALP $\phi$ is the angular direction of a complex scalar field $P$ that is charged under an approximate global $U(1)$ symmetry with a vacuum expectation value (VEV) $f/\sqrt{2}$. We assume that $|P| = r/\sqrt{2}$ has a large field value in the early universe, which can be an initial condition or driven by a negative Hubble-induced mass~\cite{Dine:1995uk}. Then higher-dimensional operators involving $P$ can be effective, and some may explicitly violate the $U(1)$ symmetry. A higher-dimensional term provides a kick to the angular direction of $P$, initiating the rotation. By cosmic expansion, $r$ decreases, and the higher-dimensional term becomes inefficient right after the initiation of rotation. $P$ continues to rotate while preserving the $U(1)$ charge $n_\theta = \dot\theta r^2$, or the angular momentum in field space.

The rotation is initially elliptic, which is a superposition of a circular motion and radial oscillations. If the radial motion continues, since it behaves as matter around the minimum of the potential, the universe is dominated by the energy of the radial motion well before the standard matter-radiation equality, spoiling the CMB predictions.  To prevent this, we require $P$ to be thermalized by particles in the thermal bath. After thermalization, the motion takes the form that minimizes the energy for a given charge, i.e., circular motion~\cite{Co:2019wyp}. The radius of the rotation continues to decrease and eventually reaches $f$, after which the rotation behaves as kination~\cite{Co:2019wyp,Co:2021lkc,Gouttenoire:2021wzu} and reduces its energy rapidly.

We first note that in this one-field model, it is difficult to achieve the required charge in Eq.~\eqref{eq:Yreq} because of inefficient thermalization. When the rotation is thermalized, the energy of the radial motion is transferred into the thermal bath at temperature denoted by $T_{\rm th}$,
\begin{equation}
\label{eq:dom}
    \frac{\pi^2}{30}g_* T_{\rm th}^4 \geq m_{P,{\rm th}}^2 r_{\rm th}^2,
\end{equation}
where $m_{P,{\rm th}}$ and $r_{\rm th}$ are the mass and amplitude of the oscillation when $P$ is thermalized. The inequality is saturated when the $P$ energy density dominates at thermalization. The charge density of the rotation $n_\theta = \dot{\theta} r^2$ is at most $m_{P,{\rm th}} r_{\rm th}^2$ at $T_{\rm th}$. From these, we obtain
\begin{equation}
    |Y_\theta| \leq \frac{3}{4}\frac{T_{\rm th}}{m_{P,{\rm th}}}.
\end{equation}

To maximize $Y_\theta$, let us consider the maximal possible $T_{\rm th}$, i.e., the most efficient thermalization mechanism. We introduce a Yukawa coupling of $P$ to a fermion $\chi$, $y P \chi^2$. The thermalization rate of $P$ via scattering with $\chi$ is $\Gamma_{\rm th}\simeq0.1 y^2T$~\cite{Mukaida:2012qn}. This, however, requires that the mass of $\chi$ is smaller than $T$, so that the abundance of $\chi$ is not exponentially suppressed. Requiring $y r <T$, the thermalization rate $\Gamma_{\rm th}$ is constrained to be
\begin{equation}
    \Gamma_{\rm th} \lesssim \frac{b T^3}{r^2},~~b\simeq0.1,
\end{equation}
where the inequality is saturated when the thermalization occurs when $y r \simeq T$. With this bound and the equality in Eq.~(\ref{eq:dom}), we obtain
\begin{align}
\label{eq:Ythetamax}
|Y_{\theta,{\rm one-field}}| < 3\times 10^5 \left( \frac{b}{0.1}\right)^{1/3} \left(\frac{10~{\rm MeV}}{m_{P,{\rm th}}}\right)^{1/3}.
\end{align}
$m_{P,{\rm th}}$ is generically larger than $m_P$ around the potential minimum, which should be above $4$ MeV 
to avoid excessive dark radiation from the $P$ thermal abundance~\cite{Co:2020dya}. 
Achieving Eq.~\eqref{eq:Yreq} in the one-field model would therefore require $c_\gamma<10^{-5}$. Although millicharged particles can produce such a coupling, their cosmological relic abundance places strong constraints~\cite{Dunsky:2018mqs,Harnik:2020ugb,Iles:2024zka,Cirelli:2024ssz}, requiring a cosmological scenario suppressing their abundance. (See~\cite{Li:2020wyl} for possible uncertainty of the constraints in~\cite{Dunsky:2018mqs}.) 

The difficulty in the one-field model can be overcome by multi-field generalization of the Kim-Nilles-Peloso mechanism~\cite{Kim:2004rp}, also called the clockwork mechanism. Following~\cite{Harigaya:2014eta,Harigaya:2014rga,Choi:2015fiu,Kaplan:2015fuy}, we introduce complex scalar fields $P_n$ ($n=0,1,\cdots k$) with global $U(1)$ charges $q_n$ and VEVs $v_n/\sqrt{2}$. (See~\cite{Choi:2014rja,Higaki:2014pja} for multi-axion models in non-linear realization.) Without loss of generality, we take $q_n>0$, $q_n \geq q_{n-1}$, and $q_0=1$. With generic potential terms consistent with the $U(1)$ symmetry,
only one field, an ALP, is massless and is identified as
\begin{equation}
    P_n = \frac{1}{\sqrt{2}}v_n e^{i q_n \theta}.
\end{equation}
The kinetic term of $\theta$ is
    $ \frac{1}{2}\sum q_n^2 v_n^2 \partial \theta  \partial \theta$, 
so the decay constant is 
\begin{equation}
    f = \sqrt{\sum q_n^2 v_n^2}.
\end{equation}

We introduce an ALP-photon coupling via
\begin{equation}
\label{eq:PEEbar}
    {\cal L} \supset -y P_0 E \bar{E} + {\rm h.c.},
\end{equation}
where $E$ and $\bar{E}$ are Weyl fermions with $U(1)_Y$ charges $-1$ and $1$, respectively. After integrating out $E$ and $\bar{E}$, we obtain $c_\gamma = 2$. The fermions that couple to $P_0$ may have different gauge charges as long as the QCD anomaly of the $U(1)$ symmetry is absent, so that no ALP mass is generated by QCD dynamics. $P_{n>0}$ may also couple with fermions in a way that electromagnetic and QCD anomalies of the $U(1)$ symmetry are absent.

We initiate the rotation of the ALP from that of $P_k$, which has the largest $U(1)$ charge. Denoting
\begin{equation}
    n_k = i \dot{P_k^*} P_k + {\rm h.c.},
\end{equation}
$Y_\theta$ is given by
\begin{equation}
    Y_\theta = q_k \frac{n_k}{s},
\end{equation}
and $n_k/s$ is constrained by Eq.~\eqref{eq:Ythetamax}. If $q_k \gg 1$,  $Y_\theta$ may be large enough to explain cosmic birefringence. In the meantime, the additional upper bound on $n_k/s$ arising from requiring the initial field value $P_k$ to remain below the Planck scale becomes weaker than Eq.~(\ref{eq:Ythetamax}).

For example, we may consider $q_n = j^n$, where $j$ is an integer, for which $q_k$ is exponentially enhanced. Such an exponentially large charge is used in~\cite{Choi:2015fiu,Kaplan:2015fuy} to obtain an exponentially large decay constant. Here we have pointed out that the same setup can exponentially enhance the charge density of a rotating field.

After the initiation of the rotation, the charge of $P_k$ is transferred to the other $P$ fields. At equilibrium, the $U(1)$ charge is distributed so that the free energy is minimized~\cite{Co:2019wyp,Domcke:2022wpb}. If $q_k v_k$ dominates the contribution to $f$, $P_k$ continues to hold most of the $U(1)$ charge. If not, when $q_k r_k$ decreases to $f$, most of the charge is transferred to the field whose contribution to $f$ is the largest.

We note that this mechanism can also suppress ALP isocurvature perturbations. The fluctuations of $\theta$ generated during inflation are $\delta \theta \sim H_{\rm inf}/(2\pi f_{\rm inf})$, where $H_{\rm inf}$ and $f_{\rm inf}$ are the Hubble scale and the decay constant during inflation~\cite{Linde:1991km}. Since $v_k$ during inflation and $q_k$ are large, the fluctuations of the ALP are suppressed.

\vspace{0.1cm}
\noindent
{\bf ALP rotation and axiogenesis.}---%
The rotating ALP can also explain the observed baryon asymmetry via axiogenesis~\cite{Co:2019wyp,Domcke:2020kcp,Co:2020xlh}. When the ALP couples to the bath, the $U(1)$ charge density in the ALP can be partially transferred to chiral asymmetries in the thermal bath, which are then converted to baryon asymmetry via the electroweak sphaleron process. The resultant baryon number divided by the entropy density is
\begin{align}
\label{eq:YB}
    Y_B = \left. \frac{45 c_B}{2\pi^2 g_*} \frac{\dot{\theta}}{T} \right|_{T=T_{\rm EW}},
\end{align}
where $c_B$ is a model-dependent parameter and $T_{\rm EW}\simeq 130~{\rm GeV}$ is the temperature when electroweak sphaleron processes decouple in the SM~\cite{DOnofrio:2014rug}.

After fixing $Y_\theta$ to explain the observed baryon asymmetry $Y_B\simeq 8.7\times 10^{-11}$, the birefringence angle is 
\begin{equation}
\label{eq:beta_baryon}
    \beta = 1.8\times 10^{-8}  \frac{-0.1}{c_B/c_\gamma}  \left(\frac{r_{\rm EW}}{f}\right)^2,
\end{equation}
where $r_{\rm EW}$ is the radius of rotation at $T=T_{\rm EW}$.

If the ALP couples only to the hypercharge gauge field, $c_B=0$ at the tree level. However, one-loop quantum corrections generate ALP-fermion couplings as 
\begin{align}
   {\cal L} \supset 6 c_\gamma \left(\frac{\alpha_Y}{4\pi}\right)^2 {\rm ln}\left(\frac{\Lambda}{T_{\rm EW}}\right)\frac{\partial_\mu \phi}{f} 
    \sum_{\psi} Y_\psi^2 \psi^\dag \bar{\sigma}^\mu \psi,
\end{align}
where $\psi$ are SM fermions and $\Lambda$ is the scale at which the ALP-hypercharge gauge field coupling is generated.
The resultant $c_B$ is
\begin{align}
\label{eq:cBmin}
    \frac{c_B}{c_\gamma} 
    \simeq -9.7\times 10^{-7}  {\rm ln}\left(\frac{\Lambda}{T_{\rm EW}}\right),
\end{align}
where we used the relation between $c_B$ and ALP-fermion couplings~\cite{Co:2020xlh}. When this minimal $c_B$ is chosen,
\begin{align}
    \beta \simeq  0.0019 \frac{1}{{\rm ln}(\Lambda/T_{\rm EW})} \left(\frac{r_{\rm EW}}{f}\right)^2.
\end{align}
An extra contribution to $c_B$ may arise from the mixing of $\bar{E}$ with an SM $SU(2)_L$-singlet charged lepton $\bar{e}$, which is necessary to let $E$ and $\bar{E}$ decay. We find that the resultant $c_B<0$; see Appendix~\ref{app:sign}. Therefore, the minimal model with $E$ and $\bar{E}$ {\it predicts} the sign of $\beta$ to be positive, which is consistent with the hint of cosmic birefringence.

If $r_{\rm EW}=f$, $\beta \lesssim 0.002$ even with the minimal $c_B$ inevitably generated by quantum corrections. If $r_{\rm EW}>f$, the hint of cosmic birefringence can be explained even if $|c_B/c_\gamma| > 10^{-6}$.  Then $\dot{\theta}$ is proportional to the mass of the radial direction of the rotating scalar field given by
\begin{align}
\label{eq:dtheta_baryon}
    m_P = \left. \frac{1}{C} |\dot{\theta}|\right|_{T=T_{\rm EW}} = 5.2~{\rm GeV} \frac{10^{-6}}{C} \frac{0.1}{|c_B|} \frac{g_{*,{\rm EW}}}{105},
\end{align}
where $g_{*,{\rm EW}} = g_*(T_{\rm EW})$, $C$ is the proportionality constant, and in the second equality we impose the requirement from the observed baryon asymmetry. In the multi-field model described previously, $C = 1/q_k\ll 1$ and $m_P= m_{P_k}$. Even for $c_B=\mathcal O(0.1)$, which is the case when the ALP has couplings to the $SU(2)$ gauge field and fermions with $c_\psi= \mathcal O(1)$, the hint of cosmic birefringence can be explained without overproducing baryon asymmetry. Also, $m_P$ can be made much above $4$ MeV to avoid excessive dark radiation. The requirement of this light radial mode motivates supersymmetric models.

Baryon asymmetry can be also (over)produced via non-perturbative production of helical hypercharge gauge fields~\cite{Giovannini:1997eg,Giovannini:1997gp,Kamada:2016eeb,Kamada:2016cnb,Co:2022kul}. We, however, find that the production is inefficient; see Appendix~\ref{app:CPI}.  

\vspace{0.1cm}
\noindent
{\bf Kination domination and gravitational waves.}---%
In supersymmetric theories, the potential of the radial direction of $P$ vanishes in supersymmetric limit and becomes non-zero via a soft supersymmetry-breaking term, which is nearly quadratic in $P$~\cite{Kim:1983ia,Moxhay:1984am}. Because the energy density of the rotation decreases as matter when $r>f$, the rotation may dominate the universe at a temperature $T_{\rm RM}$. As the radius reaches $f$ at a temperature $T_{\rm MK}$, the rotation behaves as a kination. At a temperature $T_{\rm KR}$, the universe becomes radiation dominated again. We call this scenario axion kination~\cite{Co:2019wyp,Co:2021lkc,Gouttenoire:2021wzu,Gouttenoire:2021jhk}.

The temperatures $T_{\rm KR}$, $T_{\rm MK}$, and $T_{\rm RM}$ can be obtained via
$\rho_{\theta}= \dot{\theta}^2f^2/2 = Y_\theta^2 s^2/(2f^2)= \rho_{\rm rad}$ at $T=T_{\rm KR}$, 
$Y_\theta = C m_P f^2/s$ at $T=T_{\rm MK}$, and
$\rho_\theta = C m_P Y_\theta s= \rho_{\rm rad}$ at $T=T_{\rm RM}$, respectively.
We find
\begin{align}
\label{eq:TRM}
T_{\rm RM}=& \,400~{\rm TeV}\frac{\beta}{0.005}  \left(\frac{f/c_\gamma}{2\times 10^9~{\rm GeV}}\right)^2  \frac{0.1}{|c_B/c_\gamma|}, \\
\label{eq:TMK}
T_{\rm MK} =& \,2.2~{\rm GeV} \left(\frac{0.005}{\beta}\right)^{1/3} \left(\frac{0.1}{|c_B/c_\gamma|}\right)^{1/3}  \left(\frac{80}{g_*}\right)^{1/3}, \\
\label{eq:TKR}
    T_{\rm KR} =& \, 20~{\rm MeV} \,\frac{0.005}{\beta} \;\frac{2\times 10^9~{\rm GeV}}{f/c_\gamma} \left(\frac{10}{g_*}\right)^{1/2},
\end{align}
where $g_*$ is the effective degrees of freedom at each temperature, and $m_P$ is fixed to explain the observed baryon asymmetry in obtaining $T_{\rm RM,MK}$. A period of matter and kination domination exists when $T_{\rm KR} < T_{\rm RM}$, giving
\begin{equation}
\label{eq:dom_con}
    \left|\frac{c_B}{c_\gamma} \right| < 2.5 \times 10^8 \left(\frac{\beta}{0.005}\right)^2 \left(\frac{f/c_\gamma}{10^{10}~{\rm GeV}}\right)^3.
\end{equation}

When new fermions exist in addition to $E$ and $\bar{E}$ in Eq.~\eqref{eq:PEEbar}, they may couple to $P_n$ with $q_n \gg 1$ without introducing an extra electromagnetic anomaly of $U(1)$, for which $|c_B|\gg c_\gamma$ is possible, reducing the duration of the kination-dominated era, or eliminating it. If such extra  particles do not exist, $|c_B/c_\gamma|$ cannot be larger than $\mathcal O(0.1)$; $c_B/c_\gamma \sim -10^{-6}$ is generated by quantum corrections as in Eq.~\eqref{eq:cBmin}, and $c_B=-\mathcal O(0.1) c_\gamma$ arises if $\bar{E}$ mixes with SM right-handed charged leptons~\cite{Co:2020xlh}. Therefore, in this minimal setup, $T_{\rm MK} > T_{\rm KR}$ and axion kination domination is predicted. 

We note that $T_{\rm KR} > 2.5$ MeV is required for successful Big Bang Nucleosynthesis (BBN)~\cite{Co:2021lkc}. Eq.~\eqref{eq:TKR} then places an upper bound on $f$,
\begin{equation}
\label{eq:maxf_earlyDom}
    \frac{f}{c_\gamma} \lesssim 1.6 \times 10^{10}~{\rm GeV} \left( \frac{0.005}{\beta} \right) 
\end{equation}
if axion domination occurs, namely Eq.~\eqref{eq:dom_con}. This prediction on $f$ is falsifiable by future astrophysical probes~\cite{Sisk-Reynes:2022sqd}. 

For $|c_B/c_\gamma|\simeq 10^{18\textrm{-}22}$, the rotation may instead dominate the universe after BBN and before recombination, enhancing small-scale structures~\cite{Co:2025lrd} or ameliorating the Hubble tension~\cite{Co:2024oek}. 
An era consistent with $T_{\rm KR} > 100$~eV and $T_{\rm RM} < 3$~keV~\cite{Co:2025lrd} requires
\begin{equation}
\label{eq:maxf_lateDom}
    \frac{f}{c_\gamma} \lesssim 10^{14}\,{\rm GeV} \times \min \left[ 6.1 \left( \frac{0.005}{\beta}  \right), 1.7\left( \frac{|c_B/c_\gamma|}{ 10^{20}}  \frac{0.005}{\beta}  \right)^{\frac12} \right] , 
\end{equation}
respectively. $f/c_\gamma$ has to satisfy either Eq.~\eqref{eq:maxf_earlyDom} or \eqref{eq:maxf_lateDom}.

\begin{figure}[t!]
    \centering    \includegraphics[width=\linewidth]{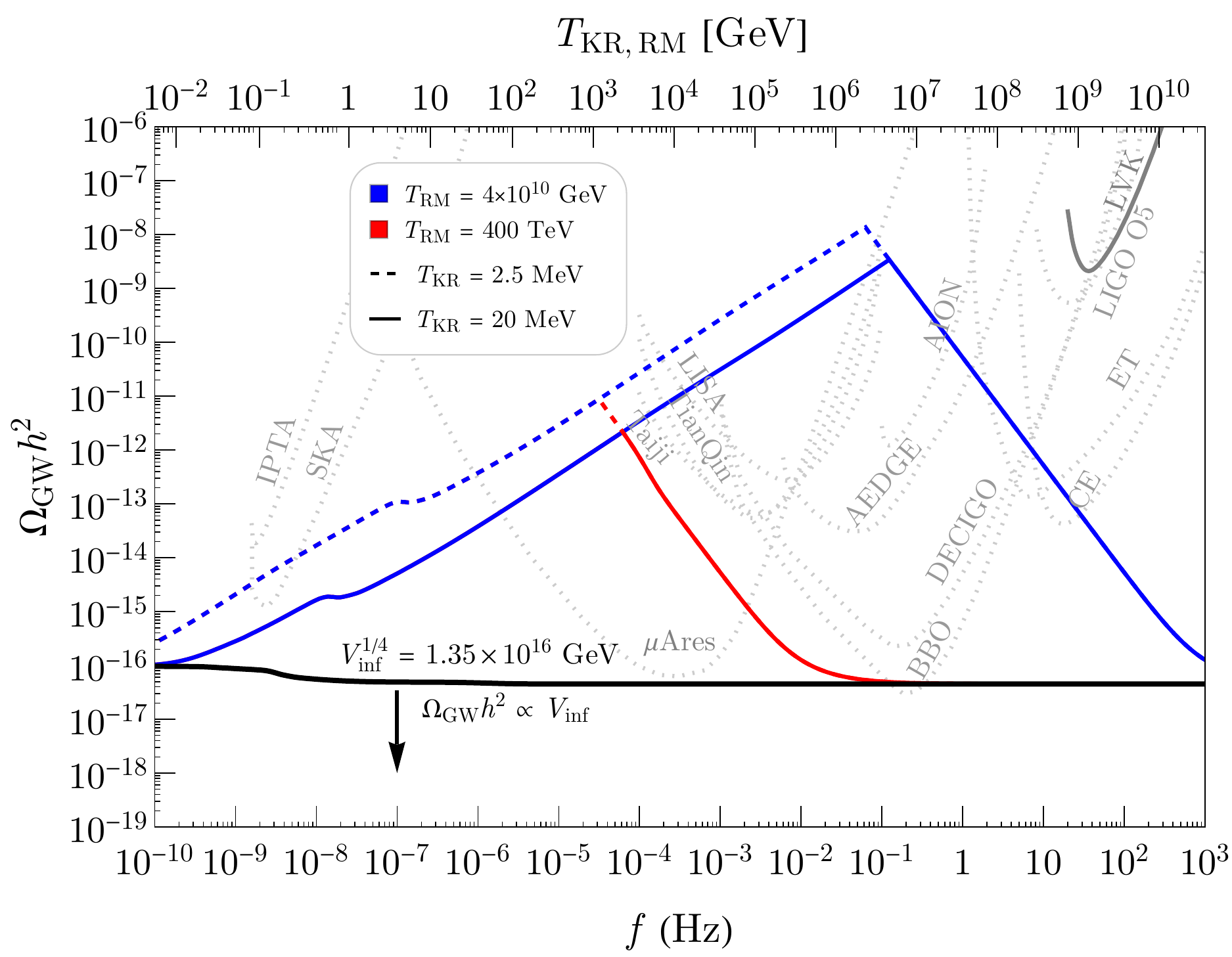}
   \includegraphics[width=\linewidth]{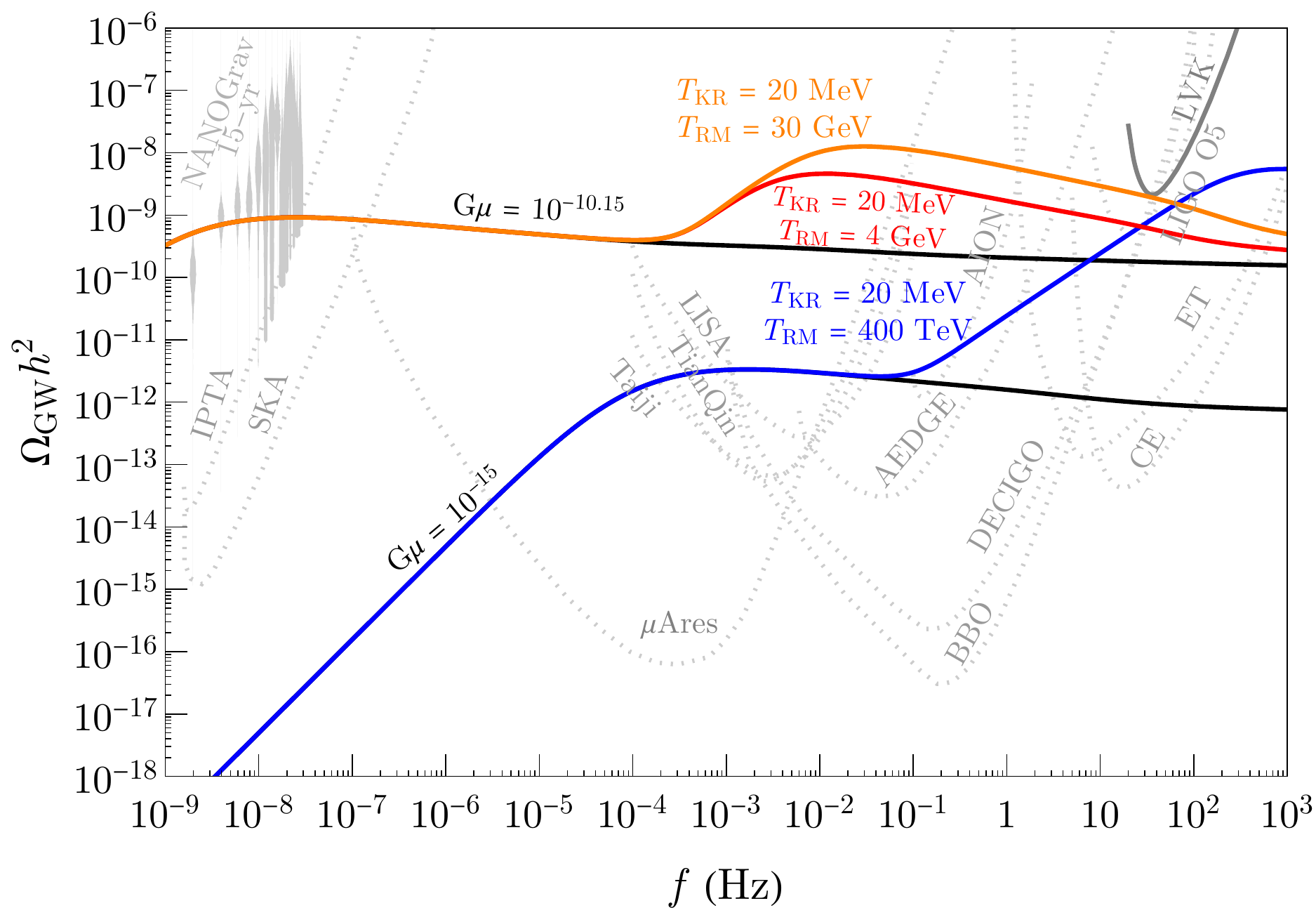}
    \caption{Spectra of gravitational waves produced by inflation (top) and local cosmic strings (bottom). Axion kination in our setup enhances gravitational waves compared to those of $\Lambda$CDM (black curves).
    Sensitivity curves are taken from~\cite{Hobbs_2010,Janssen:2014dka,Sesana:2019vho,LISA:2017pwj,Ruan:2018tsw,TianQin:2015yph,Kawamura:2020pcg,Corbin:2005ny,AEDGE:2019nxb,Badurina:2019hst,Punturo:2010zz,Reitze:2019iox,LIGOScientific:2014pky}. NANOGrav 15-year results~\cite{NANOGrav:2023hvm} and the constraint from LVK run O1-O4a~\cite{Virgo:2025aai} are also shown.}
    \label{fig:GW}
\end{figure}

Axion kination enhances gravitational waves produced by inflationary fluctuations or cosmic strings~\cite{Co:2021lkc,Gouttenoire:2021wzu,Gouttenoire:2021jhk}, as shown in the top and bottom panels of Fig.~\ref{fig:GW}, respectively. We use the approximation where the equation of state of the rotation changes from matter to kination instantaneously. (The benchmark values used in Fig.~\ref{fig:GW}  satisfy the universal bound on the duration of the kination-dominated era~\cite{Eroncel:2025bcb}.)

In the top panel, we use temperatures in Eqs.~\eqref{eq:TRM}-\eqref{eq:TKR} with $\left|c_B/c_\gamma\right|= 0.1$ and $10^{-6}$. The latter is motivated by the minimal model in Eq.~\eqref{eq:cBmin}. We take $V_{\rm inf}^{1/4}=1.35\times10^{16}\,\mathrm{GeV}$, corresponding to the limit from the combined Planck PR4, BK18, and BAO analysis, assuming standard slow-roll inflation~\cite{Tristram:2021tvh}. With $\Omega_{\rm GW}h^2 \propto V_{\rm inf}$, even for inflation scales where tensor perturbations are unobservable via the CMB, axion kination allows for detection by gravitational-wave searches.  

In the bottom panel, for $G\mu=10^{-10.15}$, which is motivated by the observations of gravitational waves by pulsar timing arrays~\cite{NANOGrav:2023hvm}, we choose the duration of each kination era such that the LIGO-Virgo-KAGRA (LVK) upper bound~\cite{Virgo:2025aai} from O1-O4a is satisfied; these choices correspond to $|c_B/c_\gamma| = 1.3\times10^3,10^4$. For $G\mu = 10^{-15}$, we choose the durations that correspond to $|c_B/c_\gamma|=0.1$. In these benchmarks, gravitational waves can be detected at both space-based and ground-based observatories.

\vspace{0.1cm}
\noindent
{\bf Summary and discussion.}---%
We have shown that the hint of cosmic birefringence and the observed baryon asymmetry can be consistently explained by the rotation of an ALP. In the minimal setup, the predicted matter- and kination-dominated eras can enhance gravitational waves produced by inflationary fluctuations or cosmic strings and make the model falsifiable through astrophysical ALP searches.

The rotation itself can produce gravitational waves. If the transition from matter to kination era is sufficiently rapid, gravitational waves are produced by so-called the Poltergeist mechanism~\cite{Inomata:2019ivs,Harigaya:2023mhl}. Also, the isocurvature perturbations of the ALP rotation, if present, can produce gravitational waves with observable anisotropy~\cite{Bodas:2025wef}. We leave the investigation of these gravitational-wave signals to future work.

Explaining the hint of cosmic birefringence requires a large angular momentum of the rotation. We proposed a mechanism to enhance the angular momentum with the aid of the Kim-Nilles-Peloso mechanism. Beyond explaining cosmic birefringence, this mechanism can also enhance the efficiency of axiogenesis, kinetic misalignment~\cite{Co:2019jts,Eroncel:2022vjg,Eroncel:2024rpe}, and acoustic misalignment mechanisms~\cite{Eroncel:2025qlk,Bodas:2025eca}. The suppressed isocurvature perturbations can be helpful in avoiding the CMB constraint on matter isocurvature perturbations in these mechanisms.

Our setup may have extra signals. For example, if the rotation's angular momentum has isocurvature perturbations, axion field perturbations grow outside the horizon during the axion kination era~\cite{Co:2020dya,Bodas:2025eca,Bodas:2025wef}. Even if the primordial fluctuations are small, CMB-scale fluctuations of the birefringence angle may be observable. Also, axion kination enhances dark matter perturbations~\cite{Redmond:2018xty,Co:2025lrd}, leaving observable imprints on small-scale structures.

In this {\it Letter}, we considered the minimal possibility where baryon asymmetry is generated by the electroweak sphaleron process. Beyond-the-SM interactions can assist the generation of baryon asymmetry from the ALP rotation, including interactions in neutrino mass models~\cite{Co:2020jtv,Chakraborty:2021fkp,Kawamura:2021xpu,Barnes:2022ren,Berbig:2023uzs,Chun:2023eqc,Barnes:2024jap,Wada:2024cbe,Datta:2024xhg,Chun:2025brc,Kuckenberg:2026oax,Datta:2026aks}, $R$-parity violation~\cite{Co:2021qgl}, and the sphaleron processes of a new gauge interaction~\cite{Harigaya:2021txz}. We leave the investigation of the implications of cosmic birefringence to those scenarios for future work.

\acknowledgments
RC and TL thank Siu Cheung Lam for his help in generating gravitational wave spectra from cosmic strings. TL thanks Sumit Biswas for fruitful discussions about CMB birefringence during TASI 2026. This work was supported by the Department of Energy grants DE-SC0025611 (RC), DE-AC02-05CH11231 (LJH), and DE-SC0009924 (KH); the NSF grant PHY-2515115 (LJH), and the World Premier International Research Center Initiative (WPI), MEXT, Japan (Kavli IPMU) (KH).

\appendix

\section{Sign of the cosmic birefringence angle}
\label{app:sign}
In this appendix, we survey a class of minimal models where at most one pair of new fermions is introduced to generate the ALP-photon coupling and show that
$c_B/c_\gamma$ is negative except for part of one model's parameter space. Therefore, $\beta$ is almost always positive after requiring the baryon asymmetry to be positive.

Since the baryon asymmetry is produced around the electroweak phase transition, where no new charged particles should exist in the thermal bath, we may use the effective field theory where all the new particles are integrated out. The relevant ALP couplings are
\begin{align}
\label{eq:LALP}
    {\cal L} =& \frac{\partial_\mu \phi}{f} \sum_{f,i}c_{f_{i}} \, f_i^\dag \bar{\sigma}^\mu f_i  \\  + & \frac{ \phi}{32\pi^2 f} \left( c_Y\, g'^{2}B^{{\mu\nu}} \tilde B_{\mu\nu}+ c_W g^{2}\; W^{\mu\nu} \tilde W_{\mu\nu}  \right), \nonumber
\end{align}
where $f_i$ are SM left-handed Weyl fermions, $B$ is the hypercharge gauge field, and $W$ is the $SU(2)_L$ gauge field. $c_\gamma = c_Y + c_W$ and $c_B$ is given by~\cite{Co:2020xlh}
\begin{align}
\label{eq:cB}
    c_B &= - \frac{12}{79} c_W \\
    &+ \sum_i\left( \frac{18}{79} c_{q_i} - \frac{21}{158}c_{\bar{u}_i} -  \frac{15}{158}c_{\bar{d}_i} + \frac{25}{237} c_{\ell_i}- \frac{11}{237}c_{\bar{e}_i} \right), \nonumber
\end{align}
where $q_i$ are quark doublets, $\bar{u}_i$ are singlet up-type quarks, $\bar{d}_i$ are singlet down-type quarks, $\ell_i$ are doublet leptons, and $\bar{e}_i$ are singlet charged leptons.

The coupling of the ALP to gluons should be absent, so that a non-zero ALP mass is not generated by QCD dynamics. The fermions responsible for the ALP-photon couplings should be color neutral.

\subsection{SM leptons}

The ALP-photon coupling may be generated by a SM charged lepton via
\begin{equation}
    {\cal L} = - \frac{P_0}{M} \ell \bar{e} H^\dag,
\end{equation}
which gives $c_Y=2$ and $c_{\bar{e}} =1$, corresponding to $c_B/c_\gamma = -11/474 <0$.

\subsection{$E$ and $\bar{E}$}

$\bar{E}$ has the same gauge charge as $\bar{e}$, while $E$ has the opposite. The possible mass terms and mixing with $\bar{e}$~are
\begin{align}
    - y P_0 E \bar{E} - m E \bar{e}: & ~~E\rightarrow E,~\bar{E}\rightarrow e^{-i\theta} \bar{E}, \label{eq:E1} \\
     - y P_0 E \bar{E} - \lambda P_0 E \bar{e}:&~~E\rightarrow e^{-i \theta}E,~\bar{E}\rightarrow  \bar{E},  \label{eq:E2} \\
      - y P_0 E \bar{E} - \lambda P_0^\dag E \bar{e}:& ~~E\rightarrow e^{i \theta}E,~\bar{E}\rightarrow  e^{-2i \theta}\bar{E},  \label{eq:E3} \\ 
    - m E \bar{E} - \lambda P_0 E \bar{e}:&~~E\rightarrow e^{-i \theta}E,~\bar{E}\rightarrow  e^{i \theta}\bar{E} \label{eq:E4}.
\end{align}
Here we also show the phase rotation of $E$ and $\bar{E}$ that removes $\theta$ from the mass terms. Eqs.~\eqref{eq:E1}, \eqref{eq:E2}, and \eqref{eq:E3} give $c_Y=2$ and $c_{\bar{E}} \geq 0$. $c_{\bar{e}}$ is generated by the mixing between $\bar{e}$ and $\bar{E}$ and is positive. The tree-level contribution is therefore $c_B/c_\gamma \leq 0$. The quantum correction to $c_B/c_\gamma$ is negative as discussed in the main text. \eqref{eq:E4} does not generate an ALP-photon coupling and is not a viable model for generating cosmic birefringence.

\subsection{$L$ and $\bar{L}$}
$L$ has the same gauge charge as $\ell$, while $\bar{L}$ has the opposite. The possible mass terms and mixing with $\ell$ are
\begin{align}
    - y P_0 L \bar{L} - m \bar{L} \ell: & ~~L\rightarrow e^{-i \theta}L,~\bar{L}\rightarrow  \bar{L}, \label{eq:L1} \\
     - y P_0 L \bar{L} - \lambda P_0 \bar{L} \ell:&~~L\rightarrow L,~\bar{L}\rightarrow  e^{-i \theta}\bar{L},  \label{eq:L2} \\
      - y P_0 L \bar{L} - \lambda P_0^\dag \bar{L} \ell:& ~~L\rightarrow e^{ -2i \theta}L,~\bar{L}\rightarrow  e^{i \theta}\bar{L},  \label{eq:L3} \\ 
    - m L \bar{L} - \lambda P_0 \bar{L} \ell :&~~L\rightarrow e^{i \theta}L,~\bar{L}\rightarrow  e^{-i \theta}\bar{L} \label{eq:L4}.
\end{align}
Here we also show the phase rotation of $L$ and $\bar{L}$ that removes $\theta$ from the mass terms. Eqs.~\eqref{eq:L1}, \eqref{eq:L2}, and \eqref{eq:L3} give $c_W=c_Y=1$.
\eqref{eq:L1} gives $c_L=1$, and positive $c_{\ell} \leq 1$ is generated by the mixing between $\ell$ and $L$ and contributes positively to $c_B$. However, from Eq.~\eqref{eq:cB}, the contribution from $c_W$ always dominates and $c_B/c_\gamma <0$. 
Eq.~\eqref{eq:L2} gives $c_L=0$ and hence $c_\ell =0$ at tree-level. Because of $c_W$, $c_B/c_\gamma <0$.
Eq.~\eqref{eq:L3} gives $c_L=2$, and positive $c_{\ell} \leq 2$ is generated by the mixing between $\ell$ and $L$ and contributes positively to $c_B$. From Eq.~\eqref{eq:cB}, when $c_{\ell} > 36/25$, $c_B/c_\gamma >0$ and $\beta$ becomes negative. This requires that  $|\lambda|^2 > 18 |y|^2/7$.
Eq.~\eqref{eq:L4} does not generate an ALP-photon coupling and is not a viable model for cosmic birefringence.

\subsection{Other electroweak representations}

Other electroweak representations may generate $c_Y$ and $c_W$, with $c_W$ having the same sign as $c_\gamma$. Because those fermions do not have mass mixing with SM fermions, $c_f=0$ at tree level.  $c_B/c_\gamma<0$ from $c_W$.

\section{Chiral Plasma Instability}
\label{app:CPI}
In this appendix, we discuss constraints from the overproduction of baryon asymmetry via non-perturbative production of helical hypercharge gauge fields. Via the ALP-hypercharge gauge field coupling and $\dot{\theta} \neq 0$, one of the helicity modes of the hypercharge gauge field experiences tachyonic instability~\cite{Turner:1987bw,Garretson:1992vt,Kamada:2019uxp,Co:2021rhi,Madge:2021abk} with a rate~\cite{Co:2022kul}
\begin{equation}
    \Gamma_{\rm CPI} \simeq \frac{\alpha_Y^2 c_5^2 \dot{\theta}^2  }{108 \pi^2 T}.
\end{equation}
where $c_5 = c_\gamma/2 \pm \mathcal{O}(10) c_B$.
When the instability develops, helical hypercharge gauge fields are produced. The helicity density is then converted into baryon asymmetry around the electroweak phase transition~\cite{Giovannini:1997eg,Giovannini:1997gp,Kamada:2016eeb,Kamada:2016cnb}, overproducing baryon asymmetry~\cite{Co:2022kul}. (See~\cite{Hamada:2025cwu,Fukuda:2025nmc}, however, for possible suppression of baryon asymmetry.) This is avoided if $\Gamma_{\rm CPI} \lesssim 20 H$ at $T_{\rm EW}$~\cite{Co:2022kul}.

For $|c_B / c_\gamma| > \mathcal{O}(10)$, after fixing $\dot{\theta}$ to explain the observed baryon asymmetry via Eq.~\eqref{eq:dtheta_baryon}, $\Gamma_{\rm CPI}$ at $T=T_{\rm EW}$ is $\mathcal{O}(10^{-20})$ GeV, which is much smaller than $20 H$ at $T_{\rm EW}$. For $|c_B/c_\gamma| \ll 1$, we obtain a lower bound on $|c_B/c_\gamma|$. Assuming that the rotation does not dominate the universe at $T_{\rm EW}$, we obtain $|c_B/c_\gamma| > 10^{-5}$ from $20 H > \Gamma_{\rm CPI}$ using Eq.~\eqref{eq:dtheta_baryon}. However, when this bound is violated, Eqs.~\eqref{eq:TRM} and \eqref{eq:TKR} show that $T_{\rm RM} > T_{\rm EW} > T_{\rm KR}$, meaning that the rotation energy actually dominates at $T_{\rm EW}$. Using $H$ self-consistently determined by the rotation energy density $\rho_{\theta}\sim \dot{\theta}^2r_{\rm EW}^2$, we find
\begin{equation}
    \left|\frac{c_B}{c_\gamma} \right| \gtrsim 10^{-8} \frac{f}{r_{\rm EW}} \frac{2\times 10^9~{\rm GeV}}{f/c_\gamma},
\end{equation}
which is satisfied even for the values of $c_B$ generated by quantum corrections given in Eq.~\eqref{eq:cBmin}, unless there is cancellation by other contributions to a percent level.

\bibliographystyle{apsrev4-2_fixed}
\bibliography{ALP}

\end{document}